\documentclass[a4paper,12pt]{article}
\usepackage{jheppub}
\usepackage{graphicx}  
\usepackage{amsmath, amssymb}   
\usepackage{dsfont} 
\usepackage{amsfonts}
\usepackage{hyperref}
\usepackage[applemac]{inputenc}
\usepackage{xcolor}
\usepackage{rotating}
\newcommand{\nn}{\nonumber \\}
\newcommand{\be}{\begin{equation}}
\newcommand{\ee}{\end{equation}}
\newcommand{\ba}{\begin{eqnarray}}
\newcommand{\ea}{\end{eqnarray}}

\newcommand{\lp}{\left(}
\newcommand{\rp}{\right)}

\newcommand{\N}{\mathcal{N}}

\newcommand{\K}{K{\"a}hler}

\newcommand{\rmi}{{\rm i}}
\newcommand{\Tr}{{\rm Tr}}

\newcommand{\purple}{\textcolor{purple}}
\definecolor{mygray}{rgb}{1, .5, .25}

\newcommand{\kt}{$K3$}
\newcommand{\cyd}[1]{$CY_{#1}$}

\newcommand{\zelltwst}{\mathcal{Z}_{K3, g}^{(ell)}}
\newcommand{\zelltw}{\mathcal{Z}_{K3, g,h}^{(ell)}}
\newcommand{\mtf}{$\mathbb{M}_{24}$}

\newcommand{\ch}{\text{ch}}
\newcommand{\slz}{$\displaystyle SL(2, \mathbb Z)$}

\title{On Mathieu moonshine and Gromov--Witten invariants} 

\author[a]{Andreas  Banlaki,}
\author[a,b]{Abhishek Chowdhury,}
\author[a]{Abhiram Kidambi,}
\author[a]{Maria Schimpf}
\affiliation[a]{Institut f\"ur Theoretische Physik, TU Wien, Wiedner Hauptstra\ss e 8-10, A-1040 Vienna, Austria}
\affiliation[b]{IIT Bhubaneshwar, SBS Building, Argul, Khordha, 752050, Odisha, India}

\emailAdd{\\banlaki@hep.itp.tuwien.ac.at} 
\emailAdd{\\achowdhury@iitbbs.ac.in}
\emailAdd{\\abhiram.kidambi@tuwien.ac.at}
\emailAdd{\\mschimpf@hep.itp.tuwien.ac.at}

\abstract{We provide further evidence that $CY_3$ manifolds are involved in an intricate way in Mathieu moonshine, i.e., their Gromov--Witten invariants are related to the expansion coefficients of the twined/twisted--twined elliptic genera of $K3$. We use the string duality between CHL orbifolds of heterotic string theory on $K3 \times T^2$ and type IIA string theory on $CY_3$ manifolds to explicitly show this connection. We then work out two concrete examples where we exactly match the expansion coefficients on both sides of the duality. 
}

\notoc

\begin{document}

\maketitle
\newpage
\tableofcontents
\section{\label{sec:intro}Introduction}
Moonshine may be labelled as a set of mysterious connections between finite sporadic groups and modular functions. For mathematicians and physicists alike, a major push in this direction came from the seminal paper by Conway and Norton on the discovery of Monstrous Moonshine \cite{CN}.  It connects two different branches of mathematics that often can be understood using the language of conformal field theory (CFT) and string theory. In mid 1980s, Frenkel, Lepowsky, and
Meurman \cite{Frenkel:1988xz} realized that the number of states at a given energy level of a 24 dimensional bosonic string theory compactified on a $\mathbb{Z}_2$ quotient of the Leech lattice \cite{leech_1967} (i.e., a chiral CFT) are dimensions of irreducible representations of the Monster group (sometimes dubbed as the ``old moonshine"). In 2010, Eguchi, Ooguri and Tachikawa (EOT) \cite{EOT} discovered a new moonshine phenomenon \cite{MirandaTwining, Gaberdiel:2010ch, Gaberdiel:2010ca, EguchiTwining, GannonModule} that connects the elliptic genus of {\kt} to the largest Mathieu group \mtf. Since the $K3$ manifold plays an important role in mathematics and physics, this new observation has led to a flurry of papers (see \cite{Duncan:2014vfa, Kachru:2016nty} for recent review articles). However, the original EOT observation has not yet been explained in a satisfactory manner, but significant progress has been made throughout the years and several new moonshine phenomena have been discovered and others have been better understood, see for example \cite{umbralone, umbraltwo, Harvey:2015mca, duncan2017onan}.

In this paper, we will be interested in strengthening a connection between Mathieu moonshine and Gromov--Witten invariants of $CY_3$ manifolds. Unlike the original Mathieu  moonshine observation which arose by considering the expansion of the $K3$ elliptic genus into $\mathcal{N}=4$ characters, the elliptic genus of $CY_3$ is rather simple when expanded in term of the $\N=2$ characters and thus one does not see anything interesting \cite{Banlaki2018}. A possible alternate route will be to look at other topological invariants such as the the Gromov--Witten invariants of the \cyd{3} to make a connection to moonshine. In particular, we will extend the idea of \cite{Cheng:2013kpa} and establish a connection between the twined and twisted--twined elliptic genera of $K3$ and Gromov--Witten invariants of certain \cyd{3} manifolds. To do this, we will rely on the string duality between compactifcations to four dimensions of heterotic $E_8\times E_8$ string theory and type IIA string theory. This allows us, on one hand, to strengthen the connection of $CY_3$ manifolds to (Mathieu) moonshine  and, on the other hand, to also establish a connection between moonshine and algebraic geometry.

As mentioned above, we will be interested in compactifications of the heterotic $E_8\times E_8$ string theory to four dimensions. In particular we will first focus on compactifications on $K3 \times T^2$. These lead to four dimensional theories with $\N=2$ spacetime supersymmetry. The action for the $\N=2$ vector multiplets is controlled by a prepotential $h(S,T,U)$, where $S$ denotes the axio--dilaton,  $T$ is the {\K} modulus and $U$ denotes the complex structure modulus of the $ T^2 $.  The prepotential $h$ is one--loop exact and it has been shown in \cite{Cheng:2013kpa} that the one--loop correction $h^{1-loop}(T,U)$ has an expansion that is connected to Mathieu moonshine. This can be related back to the EOT observation since the computation of $h^{1-loop}(T,U)$ involves the new supersymmetric index \cite{Vafa} which, for $K3$ compactifications, is related to the elliptic genus of \kt. Here we discuss compactifications in which the function $h^{1-loop}(T,U)$ not only allows an expansion involving the irreducible representations of $\mathbb M_{24}$ due to Mathieu moonshine, but is also connected to a bigger part of the {\mtf} character table through the twined and twisted--twined elliptic genera.

In particular, Chaudhuri, Hockney and Lykken (CHL) \cite{Chaudhuri:1995fk} gave a description for orbifolding the heterotic string compactified on a circle or torus in such a way that supersymmetry is not broken. While looking at such (generalized--)CHL orbifolds, one can find four dimensional $\N=2$ theories for which the one--loop corrections do not depend on the coefficients appearing in the corresponding moonshine function, but rather on the coefficients of twined functions, i.e., the \textit{McKay-Thompson} series. This has been studied explicitly for the compactifications of the heterotic $E_8 \times E_8$ string theory on $K3 \times T^2$ in \cite{Datta:2015hza, Chattopadhyaya:2016xpa, Chattopadhyaya:2017ews}. 

Compactifications of the heterotic string theory on $K3 \times T^2$ have played an important role in string--string dualities \cite{KV, FHSV}, an example being the aforementioned duality between heterotic $E_8 \times E_8$ string theory and type II string compactifications on certain \cyd{3} manifolds. These compactifications therefore give rise to the same four dimensional $\N=2$ theories. In the type IIA string theory, the one--loop threshold corrections are encoded in the Gromov--Witten invariants of the $CY_3$ manifold, while for type IIB string theory they arise from period integrals of the mirror Calabi--Yau ($CY$) manifold. Given our heterotic constructions above, we can therefore conclude that via string duality certain $CY_3$ manifolds are implicated in Mathieu moonshine, i.e., the Gromov--Witten invariants are controlled by (twisted--) characters of the elements of the Mathieu group \footnote{Following \cite{Gaberdiel:2012gf}, the generalised \textit{McKay--Thompson} series have the interpretation of being the characters of projective representations of centralizers of the group elements of $\mathbb{M}_{24}$. See (\ref{eq:projective}).}. In this paper we flesh out this general idea by providing few examples. In future we hope to provide a larger class of \cyd{3} that are implicitly involved in Mathieu moonshine via string duality.

The paper is organized as follows: In \autoref{sec:moon} we briefly review Mathieu moonshine which is then followed by  \autoref{sec:K3T2} where we discuss compactifications of the heterotic string theory on $K3 \times T^2$ and extensions to CHL orbifolds. We review the calculation of the new supersymmetric index and show its relation to Mathieu moonshine via the twined and twisted--twined elliptic genera of {\kt}. We then discuss the duality to type IIA compactifications on $CY_3$ manifolds in \autoref{sec:typeIIA}. Lastly, in \autoref{sec:examples}, we present explicit examples where we are able to match the Gromov--Witten invariants with the {expectation from the twined and twisted--twined elliptic genera on the heterotic side. We summarize our findings in \autoref{sec:conclusion}.

\section{\label{sec:moon}Mathieu moonshine}

In this section, we review some of the relevant details of the Mathieu moonshine, i.e., the connection between the {\mtf} group and the elliptic genus of the $ K3 $ surface. In general  for a superconformal field theory (SCFT) with more than $ \displaystyle \mathcal{N} = (2,2) $ supersymmetry and central charges $ \displaystyle (c , \bar c ) $  the elliptic genus is defined as \cite{Edold}
	\begin{align}
	\label{eq:zell}
	\mathcal Z^{ell} (q,y)= Tr_{RR} \left ( (-1)^{F_L + F_R} y^{J_0} q^{L_0-\frac c {24}}\bar q^{\bar{L}_0-\frac {\bar c} {24}} \right ),
	\end{align} 
	where $y:= e^{2\pi i z}, q:= e^{2\pi i \tau}$, $J_0$ is the zero mode of the third component of the affine $SU(2)$ R--symmetry algebra, $L_0$ is the zero mode of the Virasoro algebra, $F_L, F_R$ are respectively the left and right moving fermion numbers and the trace is taken over the Ramond--Ramond sector. The  right moving part is simply the  Witten index and hence for a theory with a discrete spectrum $\mathcal Z^{ell}$ is holomorphic \footnote{For  a continuous spectrum holomorphicity is not guaranteed, in fact the elliptic genus behaves like a mock modular form \cite{Troost:2010ud}. In many cases certain BPS indices exhibit wall--crossing, i.e., the value of the indices jump as one moves around in the moduli/parameter space of the theory \cite{Kontsevich:2013rda, Cecotti2010TriesteLO}.}. The elliptic genus is an index that counts BPS states of the theory and is independent of moduli of \purple{all} the theory. From a modular standpoint it is a weak Jacobi form \footnote{See \autoref{app:modularprop} for the defining properties.}  with weight $ 0 $ and index $ \displaystyle \frac{c}{6} $ w.r.t the modular group \slz. For the special case where the SCFT is an  $ \displaystyle \mathcal{N} = (4,4) $  non--linear sigma model (NLSM) on $K3$, i.e., for $c=6$, the elliptic genus is given by \cite{EOTY} (see \autoref{app:conv} for our conventions regarding Jacobi--$\theta$ functions) 
\begin{align}
\label{eq:k3ell}
\mathcal Z_{K3}^{ell} (q,y) =  8 \sum_{i = 2}^4 \left( \frac{\theta_i (q,y)}{\theta_i(q,1)} \right)^2.   
\end{align} 
In \cite{EOT} it has been noted that when (\ref{eq:k3ell}) is  expanded  in terms  of  $ \displaystyle \mathcal{N} = 4 $ superconformal characters, one finds that it  has  coefficients which are related to the dimensions of irreducible representations of the largest Mathieu group {\mtf}.
Concretely, defining the (short/long)  $ \displaystyle \mathcal{N} = 4 $ superconformal  (graded) characters in the Ramond sector as in \cite{Eguchi:2009ux} (see 
\autoref{app:N4char} for our concrete conventions)  one finds the expansion
\begin{align}
\label{eq:zellexpansion}
\mathcal Z_{K3}^{ell}(q,y) = 24\, \ch_{6,0,0} ^{\N=4}(q,y)
+ \sum_{n= 0}^\infty A_n^{(1)}\ch_{6,n,\frac 12 } ^{\N=4}(q,y),
\end{align}
where the $24 =23+1$ and other coefficients $ A_n^{(1)} $ may be related to the dimensions  of the sum of irreducible representations of {\mtf} as \footnote{Note that the character $\ch_{6,0,\frac 12}(q,y)$ here is in  a  massive representation which can be written as a sum of two  massless representations, $\lim_{h\searrow \frac 1 4} \ch^{\N=4}_{6,h-\frac 1 4,l}(q,y)= \ch^{\N=4}_{6,0,\frac 12 }(q,y)+ 2\ch^{\N=4}_{6,0,0}(q,y)$. We thank the referee for pointing this out.} \footnote{The correct decomposition of the higher  coefficients  can be fixed by requiring the twined elliptic genera to behave appropriately, see (\ref{eq:twistedellgen}) and the following discussion. }
\begin{align}
\label{eq:m24coeffs}
A_0^{(1)}&=-2= -1 -1,\nn A_1 ^{(1)} &= 90 = 45 + \overline{45},\nn   A_2^{(1)}&= 461 = 231 + \overline{231},\nn 
A_3^{(1)}&= 1540 = 770 + \overline{770}, \cdots.
\end{align}
Since the symplectic automorphisms of $K3$ form a subgroup of $\mathbb M_{23}$  \cite{Mukai} and the group of automorphisms of a NLSM on $K3$ is never  {\mtf} but in general a subgroup of the Conway group $Co_1$ \cite{Gaberdiel} the appearance of {\mtf} is not immediately clear \footnote{Progress towards a complete understanding has been made  in \cite{Gaberdiel, Cheng:2014zpa, Duncan:2015xoa, Cheng:2016org, Paquette:2017gmb}, through the concept of symmetry surfing \cite{Taormina:2011rr, Taormina:2013jza, Gaberdiel:2016iyz} and other directions  \cite{Kachru:2016ttg}.}. As the elliptic genus only counts BPS states, it is expected that similar to the case of generalized  moonshine \cite{Carnahan2008,Carnahan2009,Carnahan2012} there exists an infinite dimensional graded module \footnote{The existence of such a module has been proven in \cite{GannonModule}.}	
\begin{align}
	\label{eq:gradedmodule}
	\mathcal H^{BPS}=\bigoplus_{n=0} ^{\infty}  H_n \otimes  \mathcal H^{\mathcal N=4} _n,
	\end{align} 
	where the sum runs over the contributing irreducible $\mathcal N=4$ representations $\mathcal H^{\mathcal N=4} _n$ and $H_n$ are $\mathbb M _{24}$ representations (in general not irreducible) with $\text{dim}(H_n)=|A_n^{(1)}|$ \cite{Gaberdiel:2010ca}. This would  imply that the twined elliptic genera of $K3$ obtained by inserting an element $g\in\mathbb M_{24}$ in (\ref{eq:k3ell}), i.e.,
        \begin{align}
          \label{eq:twistedellgen}
\zelltwst (q,y) = \Tr_{RR}\left( g \,(-1)^{F_L + F_R}q^{L_0 - \frac{c}{24}} \bar{q}^{\bar{L}_0 - \frac{\bar{c}}{24}} e^{2 \pi i z J_0} \right)   
\end{align}
form an analogue of the \textit{McKay--Thompson} series of monstrous moonshine, i.e., they admit an expansion similar to (\ref{eq:zellexpansion}) but now with coefficients $\tilde A_n=Tr_{H_n}(g)$ and  transform as a Jacobi form of weight 0 and index 1 under the subgroup $\Gamma_0(N)$ of \slz. Based on these facts, in \cite{MirandaTwining, Gaberdiel:2010ch, Gaberdiel:2010ca, EguchiTwining} explicit expressions for all the twined genera  have been found. For some cases where $g\in\mathbb M_{23}$ admits an interpretation as a symplectic automorphism of $K3$ the corresponding twined genera have been calculated directly in \cite{David:2006ji}  and are in agreement with the above results. The symmetries of NLSM on $K3$ have been classified in \cite{Gaberdiel, Cheng:2016org}. Furthermore, all possible twining genera of NLSM on $K3$ have been conjectured in \cite{Cheng:2016org} based on works of \cite{Cheng:2014zpa, Duncan:2015xoa}.  This conjecture has been proven in a ``physical" way by demanding absence of unphysical wall--crossings in \cite{Paquette:2017gmb}.

In analogy to generalized moonshine, in \cite{Gaberdiel:2012gf} all the twisted (by $g$), twined (by $h$) elliptic genera $\mathcal{Z}_{K3,g,h}^{ell}$ were calculated. For every commuting pair of $g,h \in$  {\mtf}   they are defined as
\ba
\label{elliptic}
\mathcal{Z}_{K3,g,h}^{ell}(q,y)=Tr_{RR,g} \left ( h \,(-1)^{F_L} y^{J_0} q^{L_0-\frac c {24}} (-1)^{F_R}  \bar q^{L_0-\frac {\bar c} {24}} \right ),
\ea
where the trace is now taken over the $g$--twisted Ramond sector. $\mathcal{Z}_{K3,g,h}^{ell}$ are expected to fulfil certain properties. Firstly, under modular transformations they  transform as 
\ba
\label{modular}
\mathcal{Z}_{K3,g,h}^{ell}(\tau, z+ l \tau+ l^\prime) &=& e^{-2\pi i m (l^2 \tau+ 2 l z)} \mathcal{Z}_{K3,g,h}^{ell}  \:, \: l,l^\prime \in \mathbb Z \: \\ \nonumber
\mathcal{Z}_{K3,g,h}^{ell}\left(\frac {a\tau+b}{c\tau+d}, \frac z {c\tau+d} \right) &=& \chi_{g,h}\bigl(\begin{smallmatrix} a & b \\ c & d\end{smallmatrix}\bigr) e^{2\pi i \frac{c z^2}{c\tau+d}}  \mathcal{Z}_{K3,h^cg^a,h^dg^b}^{ell}(\tau,z), \: \left(\begin{matrix} a & b \\ c & d\end{matrix}\right) \in SL_2(\mathbb Z),
\ea
for a certain multiplier $\chi_{g,h}:SL(2,\mathbb Z) \rightarrow U(1)$.
The multipliers are assumed to be constant under conjugation of the pair $(g,h)$ by an element of $k\in $ {\mtf} up to a phase $\xi_{g.h}(k)$ (that  depend on certain 2--cocycles $c_g$) and to have a well defined  expansion in terms  $\mathcal N=4$ superconformal characters. In particular, $\mathcal{Z}_{K3,e,h}^{ell}$  agrees with the corresponding twined character. It is postulated in \cite{Gaberdiel:2012gf} that these properties (in particular $\chi_{g,h}, \,\xi_{g,h}$  and $c_g$) are all controlled by a 3--cocycle $\alpha$ representing a class in $H^3(\mathbb M _{24},U(1))$. Moreover, the twisted, twined and twisted--twined genera transform amongst each other following modular transformations as in  (\ref{modular}). The set of twisted--twined elliptic genera modulo modular transformations is in one to one correspondence with conjugacy classes of abelian subgroups of $\mathbb M _{24}$ generated by two elements $(g,h)$ \footnote{There are 55 such subgroups, 21 of which correspond to cyclic subgroups generated by elements of the  form $(e,g)$ and hence are associated to twining genera $\mathcal{Z}_{K3,g}^{ell}$ already discussed before in \cite{Gaberdiel:2010ch}. This also includes the twisted and twisted--twined genera obtained from the twined genera by modular transformations. The remaining 34 twisted, twined genera  were calculated in \cite{Gaberdiel:2012gf} using the  properties discussed above.}. While calculating these genera using the properties mentioned above, many of them vanish identically owing to obstructions, i.e., situations where the properties only allow for vanishing characters. Using the shorthand $ \displaystyle \zelltw := \phi_{g,h}$, the only un--obstructed twisted--twined elliptic genera (excluding twined genera) are
\begin{align}\label{eq:twtwgen}
\phi_{2B,4A_2}, \phi_{4B,4A_3}, \phi_{4B,4A_4}, \phi_{2B,8A_{1,2}}, \phi_{3A,3A_3}, \phi_{3A,3B_1},
\end{align} 
where the last two cases in the above list vanish. 

As it will be of importance for us going forward, in the special cases where $(g,h)=(\tilde g ^r, \tilde g^s)$ for $\tilde g \in \mathbb M_{24}$ where $(r,s) \in \{0,\dots,N-1\}$ and  $N=\textrm{ord}(g)$, which in particular includes the cases where the subgroups generated by $g$ and $h$  are cyclic, compact expressions for $\displaystyle \zelltw$ are known \cite{Gaberdiel:2012gf, Chattopadhyaya:2016xpa}. For such cases we define $ \mathcal Z ^{ell}_{r,s} := \mathcal Z^{ell}_{\tilde{g}^r,\tilde{g}^s}$ such that
\ba \label{twtellgen}
\mathcal Z ^{ell}_{r,s}(q,y)=\left\{
\begin{array}{ll}
\frac{N}{8}	\alpha_{\tilde{g}}^{(0,0)}  \mathcal Z^{ell}(q,y) & (r,s)=(0,0)	\\
\frac{N}{8} \alpha_{\tilde{g}}^{(r,s)} \mathcal Z^{ell}(q,y)+ N \beta_{\tilde{g}}^{(r,s)}(q)  \frac{\theta_1^2(q,y)}{\eta^6(q)} &  (r,s)\neq (0,0),
\end{array}\right.
\ea
where $ \alpha_{\tilde{g}}^{(r,s)}$ are numerical constants and $\beta_{\tilde{g}}^{(r,s)}(q)$  is a weight 2 modular form under $\Gamma_0(N)$. Explicit expressions for $\mathcal Z ^{ell}_{r,s}$ can be found in Appendix E of \cite{Chattopadhyaya:2017zul}.

\section{\label{sec:K3T2}Heterotic compactifications on $K3 \times T^2$ and CHL orbifolds}

In this section we review compactifications of $E_8\times E_8$ heterotic string theory on $K3\times T^2$ and extension to CHL orbifolds thereof. Our discussion is based on some of the results obtained in \cite{Harvey-Moore,Stieberger,Datta:2015hza,Chattopadhyaya:2017zul,Chattopadhyaya:2016xpa}. We will eventually  be interested in the prepotential of the vectormoduli and the spectrum of these theories in order to facilitate comparison to potential dual type IIA compactifications. 

\subsection{Instanton embedding and Hilbert space structure}
The fact that $K3$ has a non--vanishing curvature 2--form and as a consequence of the Bianchi identity for the threeform field strength $H ^{(3)} $, there is a requirement to embed $24$ instantons \footnote{24 is the Euler number of {\kt}.}  in $E_8\times E_8$. In general one may embed $(n^{(1)},n^{(2)})$ instantons   in some subgroups  $H_1, H_2$ of the two $E_8$'s with $n^{(1)}+n^{(2)}=24$, $n^{(i)}\in \mathbb N$ , thereby breaking the gauge group to $G_1\times G_2$, where $G_i$ is the commutant of the $H_i$ in $E_8$ \footnote{Because of symmetry we can restrict $0 \leq n_2 \leq 12$.}.  This results in a  $\mathcal{N}=(0,4)$  world--sheet supersymmetry and $\mathcal{N}=2$ supersymmetry in four spacetime dimensions.
The embedding where one chooses $n_1=24, \,n_2=0$ and $H_1=SU(2)$, $G_1=E_7$ is called the standard embedding. For the standard embedding, the first $ \displaystyle E_8 $ lattice breaks to $ \displaystyle D_6 \times D_2 $ and using the fermionic description of these lattices, one obtains the following structure \cite{Harvey-Moore, Chattopadhyaya:2016xpa} for the internal Hilbert space
\be \label{Het_Hilbert}
\mathcal H^{internal}=\mathcal H^{(6,6)}_{D_2K3} \otimes \mathcal H _{D6}^{(6,0)} \otimes \mathcal H_{E_8}^{(8,0)} \otimes \mathcal H _{T^2}^{(2,3)} ,
\ee
where the upper index labels the  left and right  central charges.  Concretely, $\mathcal H^{(6,6)}_{D_2K3}$  consists of 4 left moving bosons on $K3$ together with the 4 fermions coming from the fermionic representation of $D_2$, and a right-moving  $ \displaystyle \bar c = 6$ supersymmetric contribution. $\mathcal H _{D6}^{(6,0)} $ contains 12 fermions coming from $D_6$,  $\mathcal H_{E_8}^{(8,0)}$  is the Hilbert space of the unbroken $E_8$ and lastly, $\mathcal H _{T^2}^{(2,3)}$ is the Hilbert space of two left moving bosons on $T^2$ and a supersymmetric $c=3$ theory on $T^2$. 
The above case further generalizes to non--standard embeddings where $2(r_1+r_2)$ fermions ($r_{1,2}$ being the rank of the vectorbundle belonging to the instantons embedded in  the first/second $E_8$) from the two $E_8$'s couple to the gauge connection, and the Hilbert space structure generalizes accordingly \cite{Harrison:2013bya}. 

\subsection{The new supersymmetric index}
We now turn to the new supersymmetric index \cite{Vafa} for heterotic compactification on $K3\times T^2$. It is the central object which allows us to determine  the vector multiplet prepotential as well as the moduli dependence  of the  gravitational couplings in the effective $\mathcal N=2$ spacetime theory \cite{Harvey-Moore, Lustone, Antoniadis:1992sa, Antoniadis,Cecotti:1992vy}. For no Wilson lines on $T^2$, the new supersymmetric index for the theory on $K3\times T^2$ takes the general  form \cite{Stieberger}  
\begin{align}
\label{eqn:znewk3}
 \mathcal Z^{new}(q,\bar q) &= \frac 1 {\eta(q)^{2}}\textrm{Tr}_R ( F (-1)^F q^{L_0-\frac c {24}} \bar q^{L_0-\frac{\bar{c}}{24} } )\Big\vert_{c=(22,9)} =  Z_{K3} (q)\cdot   Z_{2,2} (q,\bar q),\nonumber \\ \nonumber
 Z_{2,2} (q,\bar q)&=\sum_{p\in \Gamma_{2,2}}  q^{\frac 12 p_L^2} \bar q ^{\frac 12 p_R^2} =\sum_{p\in \Gamma_{2,2}}  q^{\frac 12 (p_L^2-p_R^2)} e^{-2 \pi \tau_2 p_R^2},\\
 p_R ^2 &= \frac 1{2 T_2 U_2} |-m_1U+m_2+n_1T+n_2 TU|^2 , \\ \nonumber
\frac 12 p_L ^2 &= \frac 1 2 p_R ^2 + m_1n_1+m_2n_2 ,
\end{align}
where $(T, U)$  are respectively the K\"ahler and complex structure moduli of $T^2$ and the trace is taken over the Ramond sector of the internal $c=(22,9)$ CFT. In particular, one may observe that it factorizes into a holomorphic part coming from $K3$ and  the $E_8 \times E_8$ contributions, and  into a lattice sum $Z_{2,2}$. The supersymmetric index will in general depend on the topology of the manifold  and on the instanton numbers ($n^{(1)},n^{(2)}$). However, one may move in the hypermultiplet moduli space by (un--) higgsing the hypermultiplets, thereby changing the gauge groups $(G_1,G_2)$ without changing the index.  In the above expression \eqref{eqn:znewk3}, for the case of  $H_1=H_2=SU(2)$,  $Z_{K3}(q)$ takes the form 
\be\label{ZK3}
Z_{K3} (q) = 
- 2\left [ \frac {n^{(1)}}{24} \frac {E_6(q) E_4(q)}{\eta(q)^{24}}+\frac {n^{(2)}}{24} \frac {E_6(q) E_4(q)}{\eta (q)^{24}} \right ]=-2  \frac {E_6(q) E_4(q)}{\eta(q)^{24}}. \ee 
Since $\tau_2 \mathcal Z ^{new}$ is a non--holomorphic modular form of weight $-2$ \footnote{$ \mathcal Z^{new} $ appears in threshold corrections together with modular forms of weight 2.},  the form of $Z_{K3}(q)$ as in (\ref{ZK3}) is indeed determined, for arbitrary $H_1, H_2$, by its modular properties alone up to a multiplicative constant, which can be fixed to 1 by independent arguments \cite{Cheng:2013kpa}. 
As can be seen from the Hilbert space structure (\ref{Het_Hilbert}), $Z_{K3}(q)$ admits an expansion in terms of the characters of $ D6 $ and the elliptic genus of $K3$ at special values of the elliptic modulus is \cite{Harvey-Moore}  
\begin{align} \label{ZK3exp}
\eta(q)^{12} Z_{K3} (q)= &  
 \frac{1}{2}   E_4  \Big[ \left(\frac {\theta_2(q)}{\eta(q)}\right)^6 \mathcal{Z}^{ell}_{K3}(q,-1)+q^{\frac 1 4}\left(\frac {\theta_3(q)}{\eta(q)}\right)^6 \mathcal{Z}^{ell}_{K3}(q,-q^{1/2})\nonumber \\ & - q^{\frac 1 4 }\left(\frac {\theta_4(q)}{\eta(q)}\right)^6 \mathcal{Z}^{ell}_{K3}(q,q^{1/2}) \Big].
\end{align}
Here, the $E_4$ factor comes from the unbroken $E_8$, the $D6$ characters enter through the trace over the 12 free fermions coming from the broken $E_8$. The elliptic genus arises from taking a trace over the remaining 4 fermions, the 4 bosons in the left moving sector together with  the supersymmetric right moving sector
  in the $(R-,R), (NS+,R)$ and $(NS-,R)$ sectors \footnote{In our convention, $+$ stands for the insertion of $(-1)^{F_L}$ when taking the trace and $-$ stands for no such insertion.  For the right moving part the $(-1)^{F_R}$ insertion is always there so we leave away the $+$.}  \footnote{We have omitted the term proportional to $\left(\frac {\theta_1(q)}{\eta(q)}\right)^6 \mathcal{Z}^{ell}_{K3}(q,+1) $ coming from the $(R+,R)$ sector which vanishes because $\theta_1(q)=0$. One should however keep in mind that it might contribute to the twisted sectors. We thank the referee for pointing this out.}. The presence of the $ \mathcal Z_{K3}^{ell} $ terms allows us to  deduce a connection between  $\mathcal Z^{new}$ and the Mathieu group \mtf, analogous to \eqref{eq:zellexpansion}    \cite{Cheng:2013kpa}. In particular, one finds  
\ba\label{eq:new_ind_mathieuI}
-\frac{4 E_6(q)}{\eta^{12}}(q)= 24\, g_{h=1/4,l=0}(q) 
+ \sum_{n=0}^{\infty} A_n^{(1)} g_{h=n+1/4,l=1/2}(q)
\ea
where  
\ba\label{eq:new_ind_mathieuII}
g_{h=n+1/4,l}(q)&=& \left(\frac {\theta_2(q)}{\eta(q)}\right)^6  \textrm {ch}^{\mathcal N=4}_{6,n,l} (q,-1)+q^{\frac 1 4}\left(\frac {\theta_3(q)}{\eta(q)}\right)^6 \textrm{ch}^{\mathcal N=4}_{6,n,l}(q,-q^{1/2})\\ \nonumber 
&& - \;q^{\frac 1 4 }\left(\frac {\theta_4(q)}{\eta(q)}\right)^6  \textrm{ch}^{\mathcal N=4}_{6,n,l}(q,q^{1/2})
\ea 
and the coefficients $A^{(1)}_n$ as stated in \eqref{eq:m24coeffs} are the representation of the Mathieu group.
This shows the relation between the new supersymmetric index and the {\mtf} group.

\subsubsection{CHL orbifolds}
Following the discussion in \autoref{sec:moon}, the \textit{McKay--Thompson} series is an interesting object to study. To this end, we turn to the  study of  the new supersymmetric index for the case of CHL orbifolds of heterotic compactification on $K3\times T^2$ following \cite{Datta:2015hza,Chattopadhyaya:2017zul,Chattopadhyaya:2016xpa}. We consider a  NLSM on {\kt} with symmetry group $\displaystyle G < Co_1$ preserving the $\mathcal{N}=(4,4)$ supersymmetry of the model \footnote{In general, the {\kt} NLSM is moduli dependent and therefore the symmetry group also depends on which point of the {\kt} moduli space is considered.}. Such symmetry groups have been classified in \cite{Gaberdiel} (and in \cite{Cheng:2016org} for singular $K3s$) where it was shown that they correspond to groups preserving a 4--plane in the moduli space of the NLSM. We may construct a CHL orbifold by orbifolding by an element $g \in G$ together with a shift by $1/\textrm{ord}(g)$ along one  of the two  circles of $T^2$. The additional shift  guarantees  that the action is  fixed point free and thereby preserves supersymmetry (which in this case is $ \displaystyle \mathcal{N} = 2 $ spacetime supersymmetry)   \cite{Aspinwall:1995fw}.\\
The concrete orbifolds we will consider are of the type where $g$  is a symplectic automorphism of {\kt}, with $g$ being an element of one of the following {\mtf} conjugacy classes \footnote{There are 26 conjugacy classes in {\mtf} of which 16 have a representative in $ \displaystyle \mathbb{M}_{23} $. Of these 16 conjugacy classes, only 9 (as listed in \eqref{eq:m24geoconj}) can  be realized as a symplectic automorphism of {\kt}, i.e., are geometric \cite{Gaberdiel:2012um}.}
\begin{align}
  \label{eq:m24geoconj}
\text{1A, 2A, 3A, 4B, 5A, 6A, 7A, 7B, 8A}.
\end{align}
At certain  points in moduli space where {\kt} is expressible as $T^4/\mathbb{Z}_{\nu}$, $\nu \in \{2,3,4,6\}$ the new supersymmetric index has been calculated for $ \displaystyle K3 \times T^2 $ heterotic compactification and different instanton embeddings in $E_8\times E_8$ in \cite{Stieberger}.  CHL orbifolds of such models have been studied in \cite{Datta:2015hza,Chattopadhyaya:2017zul,Chattopadhyaya:2016xpa}.
For CHL orbifolds of $\textrm{ord}(g)=N$, the new supersymmetric index has the general form \cite{Chattopadhyaya:2016xpa}
 \be\label{ZnewCHL}
 \mathcal Z^{new,N} =-\frac 1 {2\eta(q)^{20}} \sum_{a,b=0} ^{\nu-1}  \sum_{r,s=0} ^{N-1} e^{-\frac {2\pi i ab}{\nu^2}} Z_{E_8}^{(a,b)}(q)\times Z_{E^\prime _8}^{(a,b)}(q) \times \frac 1 {2\nu} F(a,r,b,s;q) Z_{2,2}^{(r,s)} (q,\bar q),
 \ee
where  $Z_{E_8}^{(a,b)}(q),\, Z_{E^\prime_8}^{(a,b)}(q)$ are the partition functions of the shifted $E_8 $ lattice,  $Z_{2,2}^{(r,s)} (q,\bar q)$ is the twisted--twined partition function of $T^2$ and has the expression 
\ba
Z_{2,2}^{(r,s)}(q,\bar q) &=& \sum_{\overset{m_1,m_2,n_2 \in \mathbb Z} { n_1\in \mathbb Z + \frac r N}}   q^{\frac {p_L ^2}{2}}  \bar q ^{\frac{p_R^2}{2}} e^{2\pi i m_1 s /N}  ,
\ea
where the momenta are defined as in (\ref{eqn:znewk3}) and as before $T,U$ are respectively the K\"ahler and complex structure moduli of $T^2$. The twist by an element of order $N$ is reflected in the phase $e^{2 \pi i m_1 s /N}$ and the fractional values of $n_1$.
Furthermore,
\be
F(a,r,b,s;q):= \Tr _{R, g^{\prime a} \tilde{g}^r} \left(g^{\prime b} \tilde{g}^s  e^{i \pi F^{T^4}_R} q^{L_0} \bar q ^{\bar L_0}\right)
\ee
is the trace over $T^4$  with 4 left and 4 right moving bosons, and over right moving fermions in the twisted Ramond sector whose fermion number is $ F^{T^4}_R$.  $g^\prime$ incorporates the $\mathbb{Z}^\nu$ action \footnote{For concrete expressions, we refer the reader to \cite{Stieberger, Chattopadhyaya:2016xpa}.}.
 For example, in the case of the standard embedding, the orbifolded new supersymmetric index (\ref{ZnewCHL}) may be expanded in terms of the twisted--twined  elliptic genus of {\kt} as follows  \cite{Chattopadhyaya:2016xpa}
\ba \label{ZnewCHL_ellexp}
\mathcal Z^{new,N}(q,\bar q) &=&  \frac 1 N \sum_{r,s=0} ^{N-1}\frac {Z_{2,2}^{(r,s)} (q,\bar q )  E_4(q)}{\eta(q)^{12}} \times \Bigg[ \left(\frac {\theta_2(q)}{\eta(q)}\right)^6 \mathcal{Z}^{ell}_{K3,r,s}(q,-1) \\ \nonumber && +\,\left(\frac {\theta_3(q)}{\eta(q)}\right)^6 q^{1/4} \mathcal{Z}^{ell}_{K3,r,s}(q,-q^{\frac 12})- \left(\frac {\theta_4(q)}{\eta(q)}\right)^6   q^{1/4} \mathcal{Z}^{ell}_{K3,r,s} (q,q^{\frac 12 }) \Bigg],
\ea
where we have used the shorthand notation $\mathcal{Z}^{ell}_{K3,r,s}:= \mathcal{Z}^{ell}_{K3,\tilde{g}^r,\tilde{g}^s}$. The structure of $\mathcal Z^{new,N}(q,\bar q)$ can be understood in the same way as for the un--orbifolded case (\ref{ZK3exp}).
The appearance of the twisted--twined elliptic genera of $K3$ connect these supersymmetric CHL orbifolds to the generalized Mathieu moonshine as discussed in \autoref{sec:moon}. In particular, for the $(0,N)$ sector  one may obtain an expansion similar to (\ref{eq:new_ind_mathieuI}) and (\ref{eq:new_ind_mathieuII}) with $A_i^{(1)}$ now replaced with $A_i^{(N)}$, the coefficients of the corresponding \textit{MacKay--Thompson} series \cite{Datta:2015hza}. In our case of interest for the examples in \autoref{sec:examples}, i.e., for $\textrm{ord}(g)=N=2$, we have
\be
\mathcal{Z}^{ell}_{K3,0,1}(\tau, z)=8 \; {ch}^{\mathcal N=4}_{6,0,0}(\tau, z)+\sum_{n=0}^{\infty} A_{n}^{(2)} {ch}^{\mathcal N=4}_{6,n,\frac{1}{2}}(\tau, z), 
\ee
where the coefficient $8$ is the twisted Euler number of $K3$ and a few values of $A_n^{(2)}$ are given by
\be
A_{n}^{(2)}=-2,-6,14,-28,42,-56,86,-138, \ldots
\ee
They are the coefficients of the  \textit{MacKay-Thompson} series constructed out of the trace of an element of the $2A$ conjugacy class corresponding to the $\mathbb{Z}_2$ involution of $K3$ embedded in $\mathbb{M}_{24}$. Exactly along the lines of (\ref{eq:new_ind_mathieuI}), the $(0,1)$ sector of $\mathcal Z^{new,2}(q,\bar q)$ has an expansion in terms of $g_{h=1/4,l}(q)$ using (\ref{eq:new_ind_mathieuII}). The $(1,0)$ and $(1,1)$ sectors can be thought of as the modular transformed version of the $(0,1)$ sector. Hence, all the sectors can be related to Mathieu moonshine. In general, it is expected that the twisted--twined elliptic genus admits the following decomposition in terms of the characters of the $\mathcal{N} = 4$ superconformal algebra with central charge $c=6$
\begin{equation}
\label{eq:projective}
\mathcal{Z}^{ell}_{K3,r,s}(\tau, z)=\sum_{k=n+\frac{r}{N}\geq 0}^{\infty} \operatorname{Tr}_{\mathcal{H}_{\tilde{g}^r, k}}\left(\rho_{\tilde{g}^r, k}\left(\tilde{g}^s \right)\right) {ch}^{\mathcal N=4}_{6,h=\frac{1}{4}+k,l}(\tau, z).
\end{equation}  
Here, $l = \frac{1}{2}$ except when $h = \frac{1}{4}$ for which both $l = \frac{1}{2}$ and $l=0$ are present in the sum. The finite dimensional vector space $\mathcal{H}_{\tilde{g}^r, k}$ is the projective representation of the centralizer $C_{\mathbb{M}_{24}} (\tilde{g})$ which satisfies properties detailed in \cite{Gaberdiel:2012gf}.

In the case of a non--standard embedding it is  useful to express  $\mathcal Z^{new,N}$ in terms of the Eisenstein series. First note that  one may write $\mathcal Z^{new,N} $ in the general form
\be\label{ZnewN_1}
\mathcal Z^{new,N} (q,\bar q) = - \;4  \sum_{r,s=0} ^{N-1} Z ^{(r,s)}_{2,2}(q,\bar q) f^{(r,s)} (q),
\ee 
where $\eta (q)^{24} f^{(r,s)}(q)$ is a modular form of weight 10 under $ \displaystyle \Gamma_0(N)$ \footnote{The vector space of  such forms for $N=2$ is three dimensional and may be generated by $E_4 E_6,\, E_4 ^2 \mathcal E_2$ and $\mathcal E_2 ^2 E_6$.}.
We  start by giving the expression $\mathcal Z^{new,N}$ for the standard embedding, making use of (\ref{ZnewN_1}). Here one  obtains  \cite{Chattopadhyaya:2017zul}
\be\label{ZnewCHL_stand}
f^{(r,s)}(q) =  \frac 1 {2 \eta(q)^{24}} E_4 (q) \left[ \frac 1 4 \alpha^{(r,s)}_{\tilde{g}} E_6(q)-\beta^{(r,s)}_{\tilde{g}}(\tau) E_4(q)\right],
\ee
where $\alpha^{(r,s)}_{\tilde{g}}, \beta^{(r,s)}_{\tilde{g}}$ are the same as in (\ref{twtellgen}) and explicit expressions can be found in the Appendix E of \cite{Chattopadhyaya:2017zul}.
 For the current case of interest, i.e., $N=2$ (order 2 orbifold), we have
\ba\label{ellcoeff}
	&&\alpha^{(0,0)}_{\tilde{g}}=4\:,\: \alpha^{(r,s)}_{\tilde{g}} = \frac 4 3 \:,\: (r,s)\neq (0,0)\\ \nonumber
		&&\beta^{(0,0)}_{\tilde{g}}(\tau)=0\:,\:\beta^{(0,1)}_{\tilde{g}}(\tau)=-\frac 2 3 \mathcal E_2 (\tau) \:,\:\beta^{(1,0)}_{\tilde{g}}(\tau) = \frac 1 3 \mathcal E_2 \left(\frac{\tau} 2\right)\:,\: \beta^{(1,1)}_{\tilde{g}}(\tau) = \frac 1 3 \mathcal E_2 \left(\frac{\tau+1} 2\right)
	\ea	
where $\mathcal E_N$  is a modular form of weight 2 under $\Gamma_0(N)$ defined as	
	\be
	\mathcal E_N (\tau)= \frac {12i}{\pi (N-1)} \partial_\tau \ln \frac {\eta(\tau)}{\eta(N\tau)}.
	\ee

Generalization  of the new supersymmetric index  to  the case of non--standard embedding can be obtained from (\ref{ZnewCHL}). In particular, for a CHL orbifold of order 2, the authors of \cite{Chattopadhyaya:2016xpa} calculated $\mathcal Z^{new,2}$ at the point in {\kt} moduli space where {\kt} is expressible as $T^4/\mathbb Z_4$. The concrete expression may again be given with help of (\ref{ZnewN_1}) \cite{Chattopadhyaya:2016xpa}  \footnote{In \cite{David:2006ji}, the relevant $\alpha_{\tilde{g}}^{(r,s)}$ and $\beta_{\tilde{g}}^{(r,s)}$ for CHL orbifolds of the class $pA$ with $p=2,3,5,7$ has already been computed at the point in {\kt} moduli space where {\kt} is expressible as $T^4/\mathbb Z_2$. As an alternative, since $f^{(r,s)}(q)$ has the following transformation property under the $SL(2,\mathbb{Z})$ generators
		\ba
		f^{(r, s)}(\tau+1)&=&f^{(r, s+r)}(\tau) \\ \nonumber
		f^{(r, s)}\left(-\frac{1}{\tau}\right)&=&\left(-i \tau\right)^{-2} f^{(N-r, s)}(\tau)=(-i \tau)^{-2} f^{(r, N-s)}(\tau),
		\ea		
knowing a subset of them is enough to construct all of them. In our case of order 2 orbifolding, knowing
 $f^{(0,0)}(q)$ and $f^{(0,1)}(q)$ is enough.}:
	
\ba\label{Znew_ord2}
f^{(0,0)}(q)&=& \frac 1 {2 \eta^{24}(q)} E_4 (q)  E_6(q) \\ \nonumber
f^{(r,s)}(q)&=& \frac 3 {4 \eta ^{24}(q)} \left(
\hat b\cdot \left(\frac 3 2 \beta_{\tilde{g}}^{(r,s)} (\tau)\right)^2+\left(\frac 2 3 - \hat b \right)E_4(q)\right)\\ \nonumber &&\times\left[ \frac 1 4 \alpha^{(r,s)}_{\tilde{g}} E_6(q)-\beta^{(r,s)}_{\tilde{g}}(q) E_4(q)\right] 
\ea
The constant $\hat b$ depends on the 14 different possible instanton  embeddings listed in \cite{Stieberger} and it takes four different values:
\be \label{eq:embd_type}
\hat b \in \left\{0,\frac 4 9, \frac 2 3 ,\frac 8 9 \right\}.
\ee
The value of $\hat b$  also directly determines the difference between the number of hyper and vector multiplets through the relation
\be \label{eq:vec_hyp_diff}
N_h-N_v=144 \,\hat b -12.
\ee
As has been noted in   \cite{Chattopadhyaya:2016xpa},  an expansion in terms of the twisted--twined elliptic genera $\mathcal Z^{ell}_{K3,r,s}$ similar to (\ref{ZnewCHL_ellexp}) also exists in this case. This can be seen by comparing with (\ref{ZnewCHL_stand}) where $E_4(q)$  has been replaced by $\left(\hat b\cdot \left(\frac 3 2 \beta_{\tilde{g}}^{(r,s)} (\tau)\right)^2+\left(\frac 2 3 - \hat b \right)E_4(q)\right)$ in the twisted--twined sector but the other factor $ \left( \frac 1 4 \alpha^{(r,s)}_{\tilde{g}} E_6(q)-\beta^{(r,s)}_{\tilde{g}}(q) E_4(q)\right)$ which corresponds to the  elliptic genus of the 2A orbifold of $K3$ is however unchanged \footnote{Note that it is  $ \left( \frac 1 4 \alpha^{(r,s)}_{\tilde{g}} E_6(q)-\beta^{(r,s)}_{\tilde{g}}(q) E_4(q)\right)$ which admits a Mathieu moonshine expansion along the lines discussed after (\ref{ZnewCHL_ellexp}). We give the explicit expressions in \autoref{app:znewexp}.}.

\subsection{Threshold corrections and the prepotential}
Having obtained the new supersymmetric index, we now turn to calculating the one--loop corrections to the prepotential of the vectormoduli for the $d=4, \, \N=2$ supersymmetric effective theory arising from the CHL--orbifolded $E_8\times E_8$ heterotic string theory compactified on $K3\times T^2$ without Wilson lines \cite{Harvey-Moore, Chattopadhyaya:2017zul}. The starting point are the gravitational couplings $F_g, \, g\in \mathbb N, \, g\geq 1$ of the theory  where $g$ in the dual type IIA theory will label the genus of the Gromov--Witten invariants. They appear in the  one--loop action of the effective theory as terms of the form
\be
S^{1-loop}=\int F_g(T,U,\bar T, \bar U) F_+^{2g-2} R^2_{+},
\ee  
where $F_+$  is the anti--self dual field strength of the graviphoton and $R_+$ is the self--dual part of the Riemann curvature tensor. In our model the couplings $F_g$ depend on the (complexified) K\"ahler modulus $T$ and the complex structure $U$ of the torus $T^2$ (in  general they depend on the  vectormoduli). Compact expressions for the gravitational couplings  $F_g$ can be found in \cite{Harvey-Moore, Antoniadis:1995zn}. Though $F_g$ by itself is intricate, one is often interested in the `holomorphic part' $\bar F_g^{hol}(T,U)$ of $F_g(T,U,\bar T, \bar U)$ which can be obtained by sending $T,U \rightarrow \infty$ while  holding $\bar T, \bar U$ fixed and then taking the complex conjugate of the expression \cite{Marino:1998pg, Bershadsky:1993cx}.  It is $\bar{F}_g^{hol}$ that predicts certain topological invariants of the dual \cyd{3}, see \autoref{sec:typeIIA}. In the heterotic frame $F_g$ is given in terms of a one loop integral over the fundamental domain and one can follow the method of orbits \cite{Florakis:2016boz} to evaluate it (and also the holomorphic counterpart) for genus $g\geq1$ for several CHL--orbifoldings by conjugacy classes of $\mathbb{M}_{24}$ \cite{Chattopadhyaya:2017zul}. In particular, $\bar{F}_g^{hol}$ enjoys an expansion in polylogarithm functions ($\mathrm{Li}_{3-2g}$) and the expansion holds true for genus $g=0$. 
The genus $g=0$ holomorphic one--loop corrections to the vectormoduli prepotential of the $d=4$, $\N=2$ heterotic CHL orbifolds by a  $\mathbb{Z}_N$ automorphism acting on $K3$ together with the $\frac{1}{N}$ shift on one of the circles of $T^2$ is \cite{Chattopadhyaya:2017zul} \footnote{In \cite{Cheng:2013kpa}, the authors have shown that the vectormoduli prepotential (up to a quadratic polynomial in $S$, $T$ and $U$) of the  heterotic string theory compactified on $K3 \times T^2$ without the Wilson lines is given by
\be
\label{vec_pre}
F_{vec}=S T U+\frac{1}{3} U^{3}+\frac{1}{(2 \pi i)^{3}} c(0) \zeta(3)-\frac{2}{(2 \pi i)^{3}} \sum_{n_1>0, n_2 \in \mathbb{Z} \atop n_1=0, n_2>0} c(n_1 n_2) \textrm{Li}_{3}\left(q_{T}^{n_1} q_{U}^{n_2}\right)+\mathcal{O}\left(e^{2 \pi i S}\right)
\ee
In the weak coupling limit, i.e., when $S^{het.}\rightarrow \rmi \infty$ the 1--loop contribution is proportional to $\bar F_0^{hol,N} (T,U)$ as in (\ref{het_prepot}) for $N=1$. This correspondence also holds for the CHL orbifolds.}	
\be
\label{het_prepot}
\bar F_0^{hol,N} (T,U) = \frac {1}{\pi^2} \sum_{s=0}^{N-1} \left( \sum_{(n_1,n_2)>0} e^{-2\pi \textrm{i} n_2s/N} c^{ ({n_1}/N,s)}(n_1n_2) \textrm{Li}_{3} \left(e^{2\pi i \left(n_1 T/N+n_2 U\right)}\right) + \frac {c^{ (0,s)}(0)} 2 \zeta(3) \right)
\ee
where  the sum over $(n_1,n_2)$ is over the points
\ba
n_1,n_2\geq0, \: (n_1,n_2\neq(0,0))\\ \nonumber
n_1>0,n_2<0  \textrm{ and } n_1|n_2|\leq N,
\ea
$\zeta(n)$ is the Riemann zeta function and $c^{(r,s)}(l)$ are the fourier coefficients of $f^{(r,s)}$ defined in (\ref{ZnewN_1}), i.e., \footnote{Note that as twisted--twined elliptic genera of $K3$ control the coefficients $c^{(r,s)}(l)$, they are related to the (projective--)representations of Mathieu moonshine following the discussion after (\ref{ZnewCHL_ellexp}). }, 
\be
\label{fourier}
f^{(r,s)}(q)= \sum_{l\in \frac {\mathbb Z}N} c^{(r,s)}(l) \,q^l .
\ee
We will use (\ref{het_prepot})  to make comparisons to  the dual IIA theory.

\subsection{\label{subsec:het_spec}Spectrum of Heterotic CHL orbifolds}
The spectrum of the CHL--orbifolded heterotic compactifications on $K3\times T^2$  can be obtained by  generalizing the methods of \cite{Aldazabal:1995yw}. This was done in \cite{Chattopadhyaya:2016xpa} for CHL orbifolds of order 2 for the case when $K3$ is realized as $T^4/\mathbb Z_2$ and as $T^4/\mathbb Z_4$ for various (non--)standard embeddings. The resulting spectra are listed in tables 6 and 7--9, respectively of \cite{Chattopadhyaya:2016xpa}. We give the gauge group and matter content of the two examples most important to us in \autoref{table:Arad}. They arise from compactifications where $K3$ is realized as $T^4/\mathbb Z_4$. We denote the generator of the $\mathbb Z_4$ action by $g^\prime$.  In \autoref{table:Arad}, the first column gives the difference in number of the hypermultiplets and the vectormultiplets. The second column  states the gauge group and the shift--vector $(\gamma;\tilde \gamma) \in E_8\times E_8 $ that encodes the different instanton  embeddings. Different instanton embeddings will in general lead to different values for  $N_h-N_v$. The third column labels the twisted sectors of the $T^4/\mathbb Z_4$ realization of $K3$ from which the hypermultiplets arise. The fourth column gives the hypermultiplet content of the spectrum. In the last column we give the value of $\hat b$ according to (\ref{eq:vec_hyp_diff}).

The  gauge group can be higgsed \footnote{For a discussion of higgsing in $\mathcal N =2$ gauge theories we refer the reader to, e.g., \cite{Nazaroglu_2013}. The necessary branching rules may be found in \cite{yamatsu2015finitedimensional}. For example, we make use of the `cascade' gauge group breaking chain say $E_{6} \rightarrow S O(10) \rightarrow S U(5) \rightarrow S U(4) \rightarrow SU(3)\rightarrow SU(2) \rightarrow 0$ and $S O(14) \rightarrow S O(12) \rightarrow S O(10) \rightarrow S U(5) \rightarrow S U(4) \rightarrow S U(3) \rightarrow S U(2) \rightarrow 0$.}  by giving  VEV's to the  scalars in the hypermultiplets, thereby leaving $N_h-N_v$ unchanged. After maximally higgsing one may still give a VEV to the scalars belonging to the vectormultiplets which breaks the gauge group to $U(1)$ factors. This will  reduce $N_v$ while leaving $N_h$ unchanged. To the spectrum thus obtained, one still needs to add the four vectors that come from compactifying the ten dimensional metric $g_{\mu\nu}$ and the B--field $B_{\mu\nu}$ along the two 1--cycles of $T^2$. This gives the final number of  vectorfields and hypermultiplets.

  \begin{table}
  	\centering
 	\renewcommand{\arraystretch}{0.5}
 		\vspace{0.5cm}
 		\begin{tabular}{|c|c|c|c|c|}
 			\hline
 			& & & &\\
 			$N_h-N_v$ & Gauge group, & Twisted & 		
 			Hypermultiplets& $\hat b$ \\
 			& & & & \\
 			& Shift vector&sectors& & \\ \hline
 			& & & & \\
 			84 & $E_6\times SU(2)\times U(1)$ & ${g^\prime}^0$ &   
                        $(\mathbf{27,2;1)+(1,2;1)+(1,1;64)}$ &  \\
                        & $\times SO(14)\times U(1)$ & & $+2\mathbf{(1,1;1)}$ &$ \frac 2 3$ \\ 
                        & & ${g^\prime}^1+{g^\prime}^3$ & $6\mathbf{(1,1;1)}+4\mathbf{(1,2;1)}$ & \\
                        & $(2,1,1,0^5;2,0^7)$
                         & & $+2\mathbf{(27,1;1)}+2\mathbf{(1,1;14)}$ & \\
                         & & &  &\\
                        & & ${g^\prime}^2$ & $\mathbf{(1,2;14)}+6\mathbf{(1,2;1)}$  &\\ \hline
                        &&& &\\
                        116 & $SU(8)\times SU(2)$ & ${g^\prime}^0$ & 
                        $\mathbf{(28,2;1)+(1,1;64)}+2\mathbf{(1,1;1)}$ &\\
                         & $ \times SO(14) \times U(1)$ & ${g^\prime}^1+{g^\prime}^3$ & $4\mathbf{(8,1;1)}+2\mathbf{(8,2;1)}$ & $\frac 8 9$\\
                         & & & &\\
                       & $(3,1^5,0^2;2,0^7)$ & ${g^\prime}^2$ & $3\mathbf{(1,2;14)}+2\mathbf{(1,2;1)}$ & \\ \hline
 			
 		\end{tabular}
\caption{Completely higgsable models used in the examples in \autoref{sec:examples}.} \label{table:Arad}
\end{table}

\section{\label{sec:typeIIA}Dual type IIA compactifications on Calabi--Yau 3--folds}

In this section we briefly review type IIA compactification on \cyd{3}. Compactifying type IIA string theory on a \cyd{3}, $X$ with Hodge numbers $h_{1,1}$ (K\"ahler structure), $h_{2,1}$ (complex structure)  leads to a $d = 4,\,\N=2$ spacetime theory with one gravity multiplet, $h_{1,1}$ vectormultiplets and  $h_{2,1} + 1 $  hypermultiplets. Due to $\mathcal N=2$ spacetime supersymmetry there is no mixing between vectormultiplet and hypermultiplet moduli, i.e., the moduli space factorizes locally into $M_{vec}\times M_{hyp}$. Furthermore,  the dilaton is part of a hypermultiplet and   the vectormultiplet moduli space is exact at tree level. The vectormoduli are related to  K\"ahler/complex structure moduli in type IIA/B compactifications. For type IIA, the prepotential of the vectormoduli will, in general, receive instanton corrections. In contrast, type IIB string compactifications on a \cyd{3} leads to a vectormoduli prepotential that is exact at tree level.  This enables one to calculate an exact expression for the vectormoduli prepotential on the type IIA side in the form of an instanton expansion by relating it via the mirror map to the dual type IIB theory compactified on the mirror \cyd{3}, $ \tilde X $. Under the mirror map, the large K\"ahler structure point is mapped to the large complex structure point where the prepotential can be calculated on the type IIB side. The instanton corrections arise through holomorphic embeddings of the worldsheet into the \cyd{3}, where the worldsheet wraps  some 2--cycles in \cyd{3}. For the genus zero case, when the worldsheet is isomorphic to $S^2$, the  expression for the instanton corrections are of the form  \cite{Candelas:1990rm, Hosono:1993qy, Gopakumar:1998ii, Gopakumar:1998jq}
\be\label{IIA_prepot}
F^{inst}= \zeta(3) \frac {\chi(X)} 2 + \sum _{d_1,\dots,d_{h^{1,1}}} n^{0 }_{d_1,\dots,d_{h^{1,1}}}  \textrm{Li}_3( q_1^{d_1 } \dots q_{h^{1,1}}^{d_{h^{1,1}}}),
\ee
where $\zeta(n)$ is the Riemann zeta function, $\chi$ is the Euler character, $q_j:=e^{2\pi \rmi y_j}$ where $y^j$ labels the complexified K\"ahler moduli and  $n^{0 }_{d_1,\dots,d_{h_{1,1}}}$  denote the genus zero Gromov--Witten invariants \footnote{The are closely related to Gopakumar--Vafa invariants, see \cite{Marino:2002wa} for a review.} of multi--degree $d_j$. These Gromov--Witten invariants count in an appropriate way the number of holomorphic embeddings from a sphere into the \cyd{3}. The first term (which is a constant) arises from the constant embedding i.e. when the sphere is mapped to a point. 

The type IIA compactified on \cyd{3} is dual to the heterotic string compactified on $ \displaystyle K3 \times T^2 $ as studied in \autoref{sec:K3T2}. Calculations on the heterotic side can only be trusted when one considers the limit of weak coupling, i.e., when $S^{het.}\rightarrow \rmi \infty$ and then compare  the prepotential to the dual type IIA side, see (\ref{vec_pre}). Under the duality, $S^{het.}$ will be mapped to one of the K\"ahler moduli, say $y^i$, and hence in this limit we have
\be 
\label{IIA_prepot_inf}
(S^{het}\rightarrow i \infty ) \leftrightarrow (q_i= e^{2\pi i y^i} \rightarrow 0).
\ee
Comparing (\ref{het_prepot}) to (\ref{IIA_prepot}), we see that the two expressions agree \footnote{Since we are considering the situation with no Wilson lines  we will in general have to set  $h_{1,1}-3$ moduli to zero corresponding to $q_i=1$.} if we set 
\be
\label{eq:inst-hol}
F^{inst}= -2\pi^2\bar F^{hol,N}_0
\ee
and perform the identifications
\ba
&&\frac {\chi(X)} 2 = \sum_{s=0} ^{N-1} c^{(0,s)}(0),\\\nonumber
&&n^0_{n_1,n_2}= -2 \sum_{s=0}^{N-1} e^{- \frac {2 \pi \rmi n_2 s}  N} c^{(r,s)} (n_1 n_2).
\ea
Therefore, we see that the heterotic prepotential predicts the  Euler character \footnote{From our discussion above, the Euler character is also determined by the heterotic spectrum.} and the genus zero Gromov--Witten invariants of the \cyd{3}. In order to find the correct (matching) \cyd{3}'s, we need to calculate these invariants. In particular, by restricting to  those \cyd{3}'s that are realized as a hypersurface in a toric ambient space defined by a four dimensional reflexive polytope, one may calculate the Gromov--Witten invariants using the Mathemathica package \textit{instanton.m} \cite{Hosono:1994ax} \footnote{\textit{instanton.m}  calculates the Gromov--Witten invariants in the way outlined above, i.e., by making use of mirror symmetry. It takes the generators of the Mori cone of the toric ambient space as an input which are then used to define appropriate coordinates at the large complex structure point. These generators can be calculated using \textit{mori.x} which is a part of PALP \cite{Braun:2012vh}. In general, the toric ambient space is singular and also a  (star--)triangulation. Thus a decomposition into four simplices each having its origin as a vertex of the reflexive polytope describing it would resolve the singularities and guarantee a smooth \cyd{3} manifold. Different triangulations will correspond to different resolutions of the singularities and lead to different Mori cones. $mori.x$ can triangulate non--simplicial cones whose secondary fan  is at most three dimensional. In other cases, triangulations can be  obtained using TOPCOM \cite{Rambau:TOPCOM-ICMS:2002}. However, since the number of possible   triangulations increases rapidly with $h_{1,1}$, this is  a viable method  only when $h_{1,1}$ is small.}. Some results of these computations are discussed in the next section, \autoref{sec:examples}. 

\section{\label{sec:examples}Explicit examples}

As discussed in the previous sections, we wish to make use of the heterotic--type IIA duality to predict certain \cyd{3} used in type IIA string compactifications by matching their Gromov--Witten invariants to the coefficients of the expansion of the vectormoduli prepotential on the heterotic side. In this section we give two explicit examples in support of the duality.

On the heterotic side the models we consider are the CHL orbifolds of order 2 whose spectrum is explicitly given  in \autoref{table:Arad}. The prepotential for the order 2 CHL orbifolds can be calculated from  (\ref{het_prepot}) using (\ref{Znew_ord2}) and (\ref{eq:embd_type}). In order to find $CY_3$ for which the compactified type IIA theory matches  the heterotic side we proceed by looking for $CY_3$ manifolds which  produce the same number of vectormultiplets and hypermultiplets, i.e., with hodge numbers  $h_{1,1}=N_v-1,\,h_{2,1}=N_h-1$. As explained in \autoref{subsec:het_spec}, we maximally Higgs the spectrum of \autoref{table:Arad} and add the vector fields coming from compactification of the metric and the antisymmetric B--field thus obtaining $N_v$ number of vectorfields and $N_h$ number of hypermultiplets. For the methods explained in the previous \autoref{sec:typeIIA} to work, we have to restrict ourselves to cases where the spectrum can be completely higgsed. Then the vector fields of these theories are the ones coming from the torus compactification of the metric and the antisymmetric B--field. We find complete higgsing is possible for the  two cases given in \autoref{table:Arad} and this leads to two models with $(N_v, N_h)=(4,84)$ and  $(N_v, N_h)=(4,116) $. Given the vector and hypermultiplet content we can now proceed to search  the database by Kreuzer and Skarke \cite{CYwebsite} for appropriate \cyd{3}'s that are realized as  hypersurfaces in toric ambient space defined by a four dimensional reflexive polytope. For $(N_v,N_h)=(4,84)$ the database contains 4  possible toric \cyd{3}'s and for  $(N_v,N_h)=(4,116)$  there are 10. For each of these \cyd{3}'s (more precisely for their non--singular resolutions) we calculate the genus zero Gromov--Witten invariants. The instanton corrections to the prepotential on the type IIA side are then obtained from (\ref{IIA_prepot}).
 
Finally, we  compare the results  obtained from the heterotic and type IIA side. The heterotic prepotential (\ref{het_prepot}) is determined by the fourier coefficients of $f^{(r,s)}$ as in (\ref{fourier}). For order 2 CHL orbifolds they  are modular forms of weight 10  under $\Gamma_0(2)$ where   $(r,s) \in \{(0,0),  (0,  1),  (1,  0),  (1,1) \}$. Since as mentioned before  such forms span a three dimensional vector space, the heterotic prepotential should depend on three parameters only. Assuming string duality holds, it is sufficient to compare a small number of terms in the prepotentials to guarantee the matching of the full expressions. For  the two examples stated in  \autoref{table:Arad} we find \cyd{3}'s that give matching prepotentials.

Concretely, for the first example in \autoref{table:Arad}, we find that the \cyd{3} is described by the four dimensional reflexive polyhedron with points/vertices given by 
\ba
&&p_1=(1,0,0,0) , \: p_2=(1,2,0,0), \: p_3 =(1,2,4,0),\\ \nonumber &&p_4=(1,2,0,8),\: p_5=(-7,-6,-4,-8)
\ea
and has a triangulation which leads to the Mori cone generated by 
\be
l^1= (0, 1, 0, 0, 1, -2, 0),\: l^2 = (0, 0, 2, 0, 0, 1, 
1),\: l^3=(1, 0, 0, 1, 0, 0, -2).
\ee
After redefining 
\be
q_1 \rightarrow \sqrt {q_T}/q_U, q_2 \rightarrow q_U, 
q_3 \rightarrow  q_S,
\ee 
in (\ref{IIA_prepot}) and using (\ref{IIA_prepot_inf})  we obtain the following non--perturbative corrections to the prepotential (omitting the constant term)
\ba
\label{eq:corr1}
F^{inst} (T,U) \propto && 320\, \text{Li}_3(\sqrt{q_T})+160 \, \text{Li}_3(q_T)+320 \,\text{Li}_3(q_T^{3/2})
+160 \,\text{Li}_3(q_T^2)+320 \,\text{Li}_3(q_T^{5/2}) \nn 
&& +160 \,\text{Li}_3(q_T^3)  
+320 \,\text{Li}_3(q_T^{7/2})+4 \,\text{Li}_3(\sqrt{q_T}/q_U)+320 \,\text{Li}_3(q_U) \nn
&&+9712 \,\text{Li}_3(\sqrt{q_T} q_U)  
+143872 \,\text{Li}_3(q_T q_U)+1243176 \,\text{Li}_3(q_T^{3/2} q_U) \nn &&+8565248 \,\text{Li}_3(q_T^2 q_U)  +48219040 \,\text{Li}_3(q_T^{5/2} q_U) 
+239069184 \,\text{Li}_3(q_T^3 q_U) \nn &&+1058238684 \,\text{Li}_3(q_T^{7/2} q_U) +160 \,\text{Li}_3(q_U^2)+143872 \,\text{Li}_3(\sqrt{q_T} q_U^2) \nn &&+8493312 \,\text{Li}_3(q_T q_U^2) +239069184 \,\text{Li}_3(q_T^{3/2} q_U^2)+4301347072 \,\,\text{Li}_3(q_T^2 q_U^2) \nn 
&&+57667865600 \,\text{Li}_3(q_T^{5/2} q_U^2)+ \dots
\ea
This matches perfectly with (\ref{het_prepot}) using (\ref{Znew_ord2}) and $\hat b= \frac 2 3$ \footnote{This can easily be checked with, e.g., Mathematica.}.

Similarly, for the the second example in \autoref{table:Arad}, we find that  the \cyd{3} is  described by the reflexive polyhedron
\ba
&&p_1=(1,0,0,0) , \: p_2=(1,2,0,0), \: p_3 =(1,2,4,0),\:\nn && p_4=(1,2,0,4),\: p_5=(-3,-2,0,-4),
p_6=(-3,-2,-4,0),\: \nn &&p_7=(-3,-2,0,0), \: p_8=(-3,-2,-4,4), \: p_9 =(1,2,4,-4)
\ea
and has three different triangulations, one of which gives the Mori cone  generated by
\be
l^1=(2, 1, 0, 0, 0, 0, 1),\:  l^2= (0, -1, 0, 1, 1, 
0, -1),\: l^3=(0, -1, 1, 0, 0, 1, -1) .
\ee
As before, we redefine
\be
q_1 \rightarrow q_U,  q_2 \rightarrow q_S , q_3 \rightarrow \sqrt{q_T}/q_U
\ee
and we find that the first nonpertubative corrrections to the prepotential
for this case are (omitting the constant term)
\ba
\label{eq:corr2}
 F^{inst}(T,U)  \propto && 256 \textrm{Li}_3(\sqrt{q_T}) + 224 \textrm{Li}_3(q_T) + 256 \textrm{Li}_3(q_T^{3/2}) + 
 224 \textrm{Li}_3(q_T^2) + 256 \,\textrm{Li}_3(q_T^{5/2})\nn 
 && + 224 \,\textrm{Li}_3(q_T^3) + 
 256 \,\textrm{Li}_3(q_T^{7/2}) + 256 \,\textrm{Li}_3(q_U) + 10240 \textrm{Li}_3(\sqrt{q_T} q_U) \nn && + 141312 \,\textrm{Li}_3(q_T q_U) + 1253376 \,\textrm{Li}_3(q_T^{3/2} q_U) + 
 8529920 \,\textrm{Li}_3(q_T^2 q_U)  \nn 
 &&+ 48328704 \,\textrm{Li}_3(q_T^{5/2} q_U) + 
 238755840 \,\textrm{Li}_3(q_T^3 q_U) + 1059078144 \,\textrm{Li}_3(q_T^{7/2} q_U) \nn 
 &&+  224 \,\textrm{Li}_3(q_U^2) + 141312 \,\textrm{Li}_3(\sqrt{q_T} q_U^2) + 
 8528640 \,\textrm{Li}_3(q_T q_U^2) \nn 
 &&+ 238755840 \,\textrm{Li}_3(q_T^{3/2} q_U^2) + 4303479552 \,\textrm{Li}_3(q_T^2 q_U^2)\nn 
 && + 57655783424 \,\textrm{Li}_3(q_T^{5/2} q_U^2) +\dots
 \ea 
This agrees  with (\ref{het_prepot}) using (\ref{Znew_ord2}) and $\hat b= \frac 8 9$.

\section{\label{sec:conclusion} Conclusions}

Generalized Mathieu moonshine is a fascinating subject connecting the elliptic genus of $K3$ to the {\mtf} sporadic group. In the process of its understanding a host of connections between various topics have been established. 
In particular, a connection between Mathieu moonshine, Gromov--Witten invariants, and the periods of the holomorphic 3--form of \cyd{3} has been found in \cite{Cheng:2013kpa, Paquette:2014rma}.

Here we add to this relation by connecting generalized Mathieu moonshine to Gromov--Witten invariants of \cyd{3}. We find an explicit connection between the twisted--twined elliptic genera of $K3$ and Gromov--Witten invariants of \cyd{3}. This connection is established via the heterotic--type IIA string duality. Based on \cite{Chattopadhyaya:2016xpa, Chattopadhyaya:2017zul}, we study CHL orbifolds of $E_8\times E_8$ heterotic compactification on $K3\times T^2$. The vectormoduli prepotential in these theories can be obtained from the new supersymmetric index which admits an expansion in terms of the twisted--twined elliptic genera, which are the central objects of generalized Mathieu moonshine \cite{Gaberdiel:2012gf}. By heterotic--type IIA string duality, these theories are dual to type IIA string compactifications on some \cyd{3}. The coefficients of the non--perturbative corrections to the vectormoduli prepotential are given by the genus zero Gromov--Witten invariants, which count in an appropriate way the number of different embeddings of the sphere into the $CY_3$. We find two explicit examples for this duality for the case of CHL orbifolds of order 2. In both examples the  spectrum of the heterotic theory can be completely higgsed. In particular we only look for $CY_3$'s realized  as hypersurfaces in a toric ambient space. 

Building on the knowledge of the heterotic--type IIA duality for the non--orbifolded case \cite{KV, FHSV}, it would be interesting to see explicitly how the CHL orbifold action acts on the type IIA side. The two examples of dual \cyd{3}'s found here could be a starting point for this. Knowledge thereof would be helpful in constructing further explicit examples, and in the process strengthen the evidence of string--string dualities. Lastly, finding the dual \cyd{3}'s for other CHL orbifolds would be an interesting avenue of research and would further deepen the connections established in this paper and in \cite{Datta:2015hza,Chattopadhyaya:2016xpa,Chattopadhyaya:2017zul}.

\acknowledgments
We would like to thank Gabriel Cardoso, Aradhita Chattopadhyaya, Justin David, Albrecht Klemm, Johanna Knapp, Suresh Nampuri,  Harald Skarke and Timm Wrase for illuminating discussions. We also thank Timm Wrase for comments on the draft. We thank the referee for many helpful comments on the draft. This work is supported by the Austrian Science Fund (FWF): P 28552 and by the OeAD `Scientific \& Technological Cooperation with India' grant Project IN 27/2018 titled `Mathieu moonshine and $\N = 2$ Heterotic - Type II string duality'. The authors thank the Erwin Schr\"odinger International Institute for Mathematics and Physics for their hospitality during the workshop `Moonshine' where part of this work was done. AK thanks the Galileo Galilei Institute for Theoretical Physics, Florence for the hospitality and the INFN for partial support during the beginning of this work.
	
\appendix
\section{\label{app:conv}Conventions}
\subsection{Jacobi theta functions}
The Jacobi theta functions $ \displaystyle \theta_i(q,y), \ i = 1, \cdots ,4  $ are defined as
\begin{align}
\theta_1(q,y) &= -\rmi \sum_{n + \frac{1}{2} \in \mathbb{Z}} \left( -1 \right)^{ n - \frac{1}{2} } y^n q^{\frac{n^2}{2}} \nonumber \\ & = -\rmi q^{\frac{1}{8}} \left( y^{\frac{1}{2}} - y^{-\frac{1}{2}} \right) \prod_{n= 1}^\infty \left( 1- q^n \right) \left( 1 - yq^n \right) \left( 1 - y^{-1}q^n \right)\,,\\
\theta_2(q,y) &=\sum_{n + \frac{1}{2} \in \mathbb{Z}}  y^n q^{\frac{n^2}{2}} \nonumber \\
&=   q^{\frac{1}{8}} \left( y^{\frac{1}{2}} + y^{-\frac{1}{2}} \right) \prod_{n= 1}^\infty \left( 1- q^n \right) \left( 1 + yq^n \right) \left( 1 + y^{-1}q^n \right) \,,\\
\theta_3(q,y) &= \sum_{n \in \mathbb{Z}} y^n q^{\frac{n^2}{2}} \nonumber \\ &= \prod_{n= 1}^\infty \left( 1- q^n \right) \left( 1 + yq^{n - \frac{1}{2}} \right) \left( 1 + y^{-1}q^{n - \frac{1}{2}} \right)\,, \\
\theta_4(q,y) &= \sum_{n \in \mathbb{Z}} \left( -1 \right)^{ n } y^n q^{\frac{n^2}{2}} \nonumber \\ &= \prod_{n= 1}^\infty \left( 1- q^n \right) \left( 1 - yq^{n - \frac{1}{2}} \right) \left( 1 - y^{-1}q^{n - \frac{1}{2}} \right)\,,
\end{align}
where we have used $q=e^{2\pi \rmi \tau}$ and $y=e^{2\pi \rmi z}$. We also use the notation for the truncated Jacobi theta function  $ \displaystyle \theta_i(q) = \theta_i(q, y = 1), \ i = 1,2,3,4.$ The truncated Jacobi theta function can be used to define the Dedekind eta function as
\begin{align}
\eta(q) = \left( \frac{1}{2} \theta_2(q) \theta_3(q) \theta_4(q) \right)^{\frac{1}{3}} = q^{\frac{1}{24}} \prod_{n= 1}^\infty \left( 1 - q^n \right)\,.
\end{align}

\subsection{Eisenstein series}
The Eisenstein series of weights 4 and 6 can likewise be expressed in terms of the $\theta_i(q)$ functions
\begin{align}
E_{4}(q) &= \frac{1}{2} \sum_{i = 2}^4 \theta_i(q)^8\,, \\
E_{6}(q) &= \frac{1}{2} \lp \theta_4(q)^8 \left( \theta_2(q)^4 + \theta_3(q)^4 \right) + \theta_3(q)^8 \left( \theta_4(q)^4 - \theta_2(q)^4 \right) -\theta_2(q)^8 \left( \theta_3(q)^4 + \theta_4(q)^4 \right) \rp.
\end{align}
The  Eisenstein series have the following Fourier decompositions
\begin{align}
E_4(q) &= 1 + 240 \sum_{n = 1}^\infty\frac{n^3 q^n}{1- q^n} = 1 + 240 q + 2160 q^2 + \ldots \,,\\
E_6(q) &=1 - 504 \sum_{n = 1}^\infty \frac{n^5 q^n}{1- q^n} = 1 - 504 q - 16632 q^2 +\ldots .
\end{align}

\section{\label{app:modularprop}Modular properties of Jacobi forms and  $ \mathcal E_N (\tau) $ }

\begin{itemize}
	\item A weak Jacobi form $\phi(\tau,z)$ of weight $w$ and index $m$  defined on $\mathbb H \times \mathbb C$  is characterized by the following transformation properties
\ba
\phi\left(\frac {a\tau+b}{c\tau+d}, \frac z {c\tau+d} \right) &=& (c\tau +d)^w e^{2\pi i m \frac{c z^2}{c\tau+d}}  \phi(\tau,z)  \:, \: \left(\begin{matrix} a & b \\ c & d\end{matrix}\right) \in SL_2(\mathbb Z),\\
\phi(\tau, z+ l \tau+ l^\prime) &=& e^{-2\pi i m (l^2 \tau+ 2 l z)} \phi(\tau,z), \: l,l^\prime \in \mathbb Z \: .
\ea

\item $\mathcal E_N$  is defined by
\be
\mathcal E_N (\tau)= \frac {12i}{\pi (N-1)} \partial_\tau \ln \frac {\eta(\tau)}{\eta(N\tau)}.
\ee
It is a modular form of weight 2 under $\Gamma_0(N)$. It also has the following transformation properties \cite{Datta:2015hza}
\be
\mathcal E_N(\tau+1)=\mathcal E_N(\tau)\:, \: \mathcal E_N(-1/\tau)=-\tau^2 \frac 1 N \mathcal E_N(\tau/N).
\ee
Further for $N=2$, the following can be deduced \cite{Datta:2015hza}
\be
\mathcal E_2\left(-\frac 1 {2\tau} \right)=-2 \tau^2 \mathcal E_2(\tau)\:,\: \mathcal E_2\left(-\frac 1{2\tau} +\frac 1 2 \right)=\tau^2 \mathcal E_2 \left( \frac {\tau + 1} 2\right).
\ee
\end{itemize}
\section{\label{app:N4char}$\mathcal N=4$ superconformal characters}
The (graded) characters of  $\N=4$ superconformal algebra with central charge $ c $ in the Ramond sector are defined as 
\be
\ch_{c,h-\frac c {24},\ell} ^{\N=4}(q,y) = {\rm tr}_{\mathcal{H}_{h,\ell}} \left((-1)^F q^{L_0-\frac c {24}} e^{4 \pi \rmi z J_0}\right),
\ee
where  $h$ and  $\ell$ are  respectively the eigenvalues of $L_0$ (conformal weight)   and  $J_0$ (iso--spin), of the highest weight state belonging to the representation $\mathcal H_{h,\ell}$.
Unitarity requires $h\geq \frac  c {24}$.\\
The characters \cite{Eguchi:1987wf} are given by (using conventions from \cite{Eguchi:2009ux})
\begin{itemize}
	\item Massless representation exist for $h=\frac  c {24},\: \ell=0,\frac 12, \dots,\frac c{12}  $ and are given by
	\be
	\ch_{c,0,\ell} ^{\N=4}(q,y)=\frac {\rmi} {\theta_{1}(\tau,2z)} \frac{\theta_{1}(\tau,z)^2}{\eta(\tau)^3}\sum_{\varepsilon=\pm 1}\sum_{m\in \mathbb Z}
	\varepsilon \frac {e^{4\pi \rmi \varepsilon ((\frac c 6 +1)m+\ell)(z+\frac 12)}}{(1-y^{-\varepsilon} q^{-m})^2} q^{(\frac c 6+1)m^2+2\ell m}.
	\ee
	In particular for $\ell=0$ this may be written as
	\be
	\ch_{c,0,0} ^{\N=4}(q,y) = \frac {-\rmi} {\theta_{1}(\tau,2z)} \frac{\theta_{1}(\tau,z)^2}{\eta(\tau)^3} \sum_{m \in\mathbb Z}  q^{\lp\frac c {6} +1\rp m^2} y^{(\frac c 3+2) m}\frac{1+y q^m}{1- y q^m}.
	\ee
	\item Massive representation exist for $h> \frac c{24} , \:\ell=\frac 12,1,\dots,\frac c {12}$  and are given by
	\be
	\ch_{c,h-\frac c {24},\ell}^{\N=4}(q,y)=\rmi q^{h-\frac {6 \ell^2}{ c +6}-\frac c {24} }\frac{\theta_1(\tau,z)^2}{\theta_1(\tau, 2z) \eta(\tau)^3} \left(\vartheta_{\frac c 6 +1,2\ell}\left(\tau,z+\frac 12\right)-\vartheta _{\frac c 6+1,-2\ell}\left(\tau,z+\frac 12\right)\right),
	\ee
	where
	\be
	\vartheta _{P,a}(q,y)=\sum_{n\in \mathbb Z} q^{\frac{(2Pn+a)^2}{4P}} y^{2Pn+a}\,.
	\ee
\end{itemize}
At the unitarity bound $h=\frac c {24}$ the massive $\N=4$ characters decompose into massless ones according to
\ba
\label{massive_massless}
\lim_{h\searrow \frac c {24}} \ch^{\N=4}_{c,h-\frac c{24},l}(q,y)= \ch^{\N=4}_{c,0,l}(q,y)+ 2\ch^{\N=4}_{c,0,l-\frac 12}(q,y)+\ch^{\N=4}_{c,0,l-1}(q,y).
\ea 
\section{\label{app:znewexp}Explicit expansion of $\mathcal Z^{new,2}$}
In this appendix we give the explicit relationships between  $f^{(r,s)}$, defined in (\ref{ZnewN_1}),and the twisted--twined elliptic genus of $K3$ for the 2A orbifold appearing in our discussion. To make expressions more readable we only write the parts of $ f^{(r,s)}$ that are relevant for this connection.
 
By comparing (\ref{ZnewN_1}), (\ref{ZnewCHL_stand}) to (\ref{ZnewCHL_ellexp}) we get the following expansions in terms of the twisted--twined  elliptic genera $\mathcal Z^{ell}_{(r,s)}$\\
$\underline{r=s=0}:$
\ba
&&\frac {-4}{\eta(q)^{12}} E_6(q)=\Bigg[ \left(\frac {\theta_2(q)}{\eta(q)}\right)^6 \mathcal{Z}^{ell}_{K3,0,0}(q,-1) \\ \nonumber &+&\left(\frac {\theta_3(q)}{\eta(q)}\right)^6 q^{1/4} \mathcal{Z}^{ell}_{K3,0,0}(q,-q^{\frac 12})- \left(\frac {\theta_4(q)}{\eta(q)}\right)^6   q^{1/4} \mathcal{Z}^{ell}_{K3,0,0} (q,q^{\frac 12 }) \Bigg]
\ea 
$\underline{r=0,s=1:}$
\ba
&&\frac {-4}{\eta(q)^{12}}\frac 1 3 \left[ E_6(q) +  2  \mathcal E _2(\tau) E_4(q) \right]=\Bigg[ \left(\frac {\theta_2(q)}{\eta(q)}\right)^6 \mathcal{Z}^{ell}_{K3,0,1}(q,-1) \\ \nonumber &+&\left(\frac {\theta_3(q)}{\eta(q)}\right)^6 q^{1/4} \mathcal{Z}^{ell}_{K3,0,1}(q,-q^{\frac 12})- \left(\frac {\theta_4(q)}{\eta(q)}\right)^6   q^{1/4} \mathcal{Z}^{ell}_{K3,0,1} (q,q^{\frac 12 }) \Bigg]
\ea
$\underline{r=1,s=0:}$
\ba
&&\frac {-4}{\eta(q)^{12}} \frac 1 3\left[  E_6(q) - \mathcal E _2\left(\frac \tau 2\right) E_4(q) \right]=\Bigg[ \left(\frac {\theta_2(q)}{\eta(q)}\right)^6 \mathcal{Z}^{ell}_{K3,1,0}(q,-1) \\ \nonumber &+&\left(\frac {\theta_3(q)}{\eta(q)}\right)^6 q^{1/4} \mathcal{Z}^{ell}_{K3,1,0}(q,-q^{\frac 12})- \left(\frac {\theta_4(q)}{\eta(q)}\right)^6   q^{1/4} \mathcal{Z}^{ell}_{K3,1,0} (q,q^{\frac 12 }) \Bigg]
\ea
$\underline{r=s=1:}$
\ba
&&\frac {-4}{\eta(q)^{12}} \frac 1 3\left[ E_6(q) -  \mathcal E_2 \left(\frac{\tau+1}2\right) E_4(q) \right]=\Bigg[ \left(\frac {\theta_2(q)}{\eta(q)}\right)^6 \mathcal{Z}^{ell}_{K3,1,1}(q,-1) \\ \nonumber &+&\left(\frac {\theta_3(q)}{\eta(q)}\right)^6 q^{1/4} \mathcal{Z}^{ell}_{K3,1,1}(q,-q^{\frac 12})- \left(\frac {\theta_4(q)}{\eta(q)}\right)^6   q^{1/4} \mathcal{Z}^{ell}_{K3,1,1} (q,q^{\frac 12 }) \Bigg]
\ea
As these are the expressions that show up in $\mathcal Z ^{new}$ for both the standard and the non--standard embedding they  connect $\mathcal Z^{new}$ to Mathieu moonshine.

	
\bibliographystyle{JHEP}
\bibliography{refs}
	
\end{document}